\documentclass[prl,twocolumn,preprint,a4paper,nofootbib]{revtex4-1}

\usepackage{graphics, graphicx, caption, subcaption, tabularx, color, amsmath, amssymb, verbatim, epsfig, setspace, mciteplus, multirow, mhchem}
\usepackage{epstopdf}

\newcounter{chem}
\newcounter{temp}

\begin{document}

\title{Multi-Layer Free Energy Perturbation}

\author{Ying-Chih Chiang}
\author{Frank Otto}

\affiliation{Department of Physics, 
 Chinese University of Hong Kong, Shatin, N.T., Hong Kong}

\date{\today}

\begin{abstract}
Free energy perturbation (FEP) is 
frequently used to evaluate the free energy change of
a biological process, e.g. the drug binding free energy 
or the ligand solvation free energy.
Due to the sampling inefficiency, FEP is often
employed together with computationally expensive
enhanced sampling methods.
Here we show that this sampling inefficiency, which stems from 
not accounting for the environmental reorganization, is 
an inherent property of the single-ensemble ansatz of FEP, 
and hence simply prolonging the MD simulation can hardly 
alleviate the problem. 
Instead, we propose a new, multi-ensemble ansatz -- 
the multi-layer free energy perturbation (MLFEP), 
which allows environmental reorganization processes 
(relaxation) to occur automatically during the 
MD sampling. Our study paves the way toward a fast 
but accurate free energy calculation that can be 
employed in computer-aided drug design.
\end{abstract}

\maketitle

Accurately evaluating the free energy change of a ligand 
binding to its receptor has a very practical use in 
computational drug design, i.e. determining the relative 
binding free energies between two drug candidates for lead-
optimization. One of the most frequently employed method for 
this purpose is the so-called free energy perturbation (FEP) 
method~\cite{Zwanzig54}, 
which states that the free energy change between the final 
target state T and the initial reference state R can be 
evaluated via a single ensemble average, i.e. 
\begin{eqnarray}
\label{eq:FEP}
e^{-\beta \Delta A} = \langle e^{-\beta u} \rangle_{\text{R}} \;,
\end{eqnarray}
where $\Delta A$ denotes the free energy change and 
$\beta=1/k_{\text{B}}T$, with the Boltzmann factor 
denoted by $k_{\text{B}}$ and temperature by $T$.
The term $u$ denotes the perturbation introduced
to the initial reference state, and its value is
given by the potential difference between the 
target state T and the reference state, 
$u=U_{\text{T}}-U_{\text{R}}$. 
Finally, the symbol $\langle \cdots \rangle_{\text{R}}$
represents that the canonical ensemble average is performed 
over the reference state R. In other words, the sampling is
performed using the Hamiltonian of the reference state.
Similarly, one can also sample the target state T for 
calculating $\Delta A$, this leads to the so-called backward
FEP calculation, i.e. $e^{\beta \Delta A} = \langle e^{\beta u} \rangle_{\text{T}} $.
Although Eq.~\ref{eq:FEP} is theoretically exact, numerically
evaluating the ensemble average often suffers from a problem
of the sampling inefficiency.
While plenty of methods, e.g. stratification (multi-step FEP)~\cite{mFEP,POHO2010}, 
confine-and-release method~\cite{Karplus03,Roux05,Mobley07_confandrel}, 
or replica-exchange molecular dynamics (with solute tempering)~\cite{REMD,SUGI2000,Liu05,Wang11,Moors11,Friesner12}, 
have been developed to improve the sampling efficiency and 
hence advance the FEP convergence, the current computational 
cost of using enhanced sampling methods combined with FEP 
is still rather prohibitive to be regularly applied in drug design~\cite{wang2015accurate,Nathan16}. Hence, further pursuing an accurate but fast free energy method is still desirable. 

Previously we have shown that the insufficient sampling 
comes from missing the environmental reorganization~\cite{chiang16_vssangle}, 
e.g. allowing the water to move or reorient to accommodate the
inserted ligand (perturbation). This process is a type of
\emph{relaxation process}, which is well studied in
gas phase reactions. For instance, consider the quantum nuclear 
dynamics~\cite{Elke96,Simona04,Chiang11} during the 
interatomic/intermolecular  
Coulombic decay process (ICD)~\cite{Lenz97,Nico10_1,Till10,Uwe10,Kirill14,Vasili16},
in the neon dimer~\cite{Jahnke04,Simona04}. 
After introducing a strong perturbation 
to the system (ionizing an inner 
valence electron on Ne), the system quickly responds to 
this perturbation by emitting one electron on the 
neighboring Ne, resulting in a Ne$^+$-Ne$^+$ state 
that undergoes Coulomb explosion to lower the system 
(free) energy. 
Clearly, the nuclear motion in the electronic decay process
is always governed by the corresponding Hamiltonian 
of a specific electronic state~\cite{Elke96}.
Similarly, in the classical molecular dynamics, the 
molecular motion is also governed by the Hamiltonian 
of the simulated system.
The only difference is that the classical system is 
described by Newtonian mechanics~\cite{NAMD} with force fields~\cite{CGenFF}.

\begin{figure}
\centering
\includegraphics[width=0.4\textwidth]{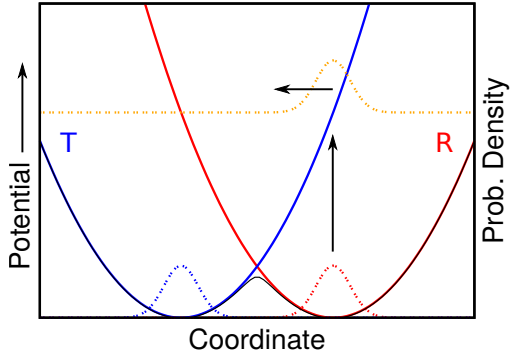}
\caption{\small Local potential trap and the insufficient 
sampling problem. The reference state R and the target 
state T have their own potential landscapes (solid curves) 
that confine the sampled probability distributions (dotted 
curves). When any conformation that belongs to the distribution
of R is placed on the landscape of T, it will move according to 
the Hamiltonian of T, leading to the reorganization process.}
\label{fig:Doublewells}
\end{figure}

Let us now consider a common illustrative example
in free energy calculations, namely, 
the ligand solvation process. 
According to Eq.~\ref{eq:FEP}, collecting the ensemble
governed by the Hamiltonian of reference state R 
(ligand and water solvent are separated) is 
sufficient for correctly evaluating $\Delta A$.
However, as illustrated in Fig.~\ref{fig:Doublewells},
the two end states can have very different potential energy 
landscapes so that their associated probability distributions
center at different geometries, as indicated by the dotted 
curves in Fig.~\ref{fig:Doublewells}. 
Consequently, when sampling the distribution via MD simulation
in order to sample all possible conformations of the reference 
state R, one faces the sampling inefficiency because 
the relevant microstates belonging to the target state T are 
generally missed, leading to a non-converged free energy
result.
This problem can be solved by introducing 
the reorganization process (relaxation) into the sampling 
procedure by starting at the same conformation as reference 
state R but performing the MD simulation based on the 
Hamiltonian of target state T,
see e.g. the orange dotted curve in Fig.~\ref{fig:Doublewells}.
While this idea may not be so familiar to 
the native biophysics society, 
its quantum version is regularly performed 
in studying gas phase molecular dynamics involving multiple electronic states~\cite{Elke96,Simona04,Chiang11,Horst84,Horstbook}.
Furthermore, our approach is very different from 
contemporary enhanced sampling schemes, e.g.
increasing temperature to overcome the potential barrier, 
adding a biasing potential to flatten the potential 
landscape, or even using the ``adiabatic" potential (black curve) for sampling~\cite{Clara07}.
These schemes focus on forcing the MD sampling to explore
a larger conformational space but continue using Eq.~\ref{eq:FEP} to evaluate $\Delta A$.
Rather, we believe that the insufficient sampling
problem is an inherent property 
such that the best way to solve it is to use a different working equation than Eq.~\ref{eq:FEP}. 

Does such a new equation, which allows the system to relax 
automatically during the simulation, exist? 
Exploiting the fact that 
$e^{-\beta \Delta A}$ is a constant under the given NVT 
ensemble, further imposing one additional sampling over a 
normalized distribution will not 
change its value, e.g. $\langle e^{-\beta \Delta A} \rangle_{\text{T}}=e^{-\beta \Delta A}$, as long as the 
sampling is sufficient. 
Hence we have,
\begin{eqnarray}
\label{eq:MLFEP_backward}
e^{-\beta \Delta A} = \langle \langle e^{-\beta u} \rangle_{\text{R}} \rangle_{T} \;,
\end{eqnarray}
where the definitions of all symbols are identical 
with Eq.~\ref{eq:FEP}.
In Eq.~\ref{eq:MLFEP_backward}, we further imposed the
sampling over the distribution of target state T,
which does not affect $\Delta A$, since its value
is already determined at the sampling of the reference
state R. While Eq.~\ref{eq:MLFEP_backward} seems 
to introduce more effort in MD sampling to 
evaluate $\Delta A$, this equation actually allows 
the environmental reorganization. Let us explain.
When evaluating Eq.~\ref{eq:MLFEP_backward}, one
first performs a short equilibrium sampling to 
collect the microstates that belongs to state T,
and then from each microstate (each frame of the collected
trajectory) one performs an MD sampling using the
Hamiltonian of state R to evaluate the free energy 
change within this simulation. Thus, each microstate
of state T gives one $e^{-\beta \Delta A}$ that will 
later participate in the ensemble average over state T.
Interestingly, for each microstate, the sampling now
always begins at a non-equilibrium high energy 
conformation. This conformation will then undergo
a relaxation process automatically due to the
governing Hamiltonian, and hence the sampling is 
more efficient than waiting for rare events to happen.
For practical purposes, Eq.~\ref{eq:MLFEP_backward} 
can also be expressed in a reversed sampling form that reads,
\begin{eqnarray}
\label{eq:MLFEP}
e^{\beta \Delta A} = \langle e^{\beta \Delta A} \rangle_{\text{R}} = \langle \langle e^{\beta u} \rangle_{\text{T}} \rangle_{R} \;.
\end{eqnarray}
This new format describes the process in Fig.~\ref{fig:Doublewells}: start the sampling under
the reference state R, and then introduce the
environmental reorganization via the relaxation process 
governed by the target state T.
One additional advantage of Eq.~\ref{eq:MLFEP} is that
we can now assign a common reference state R and save
the trajectory for evaluating $\Delta A$ between the
reference state and different target states. 
This can further save some computational effort.  
Finally, since Eqs.~\ref{eq:MLFEP_backward}-\ref{eq:MLFEP} already go beyond the usual FEP theory,
we will term our new approach as
the multi-layer free energy perturbation (MLFEP),
in order to distinguish it from the virtual substitution scan (VSS)~\cite{chiang16_vss,chiang16_vssangle} 
which is purely based on a single-ensemble approach but
also has a dual sampling format.

%

\end{document}